\baselineskip=14pt

\magnification=\magstep1

\font\msbm=msbm10
\font\bfv=cmbx10 scaled \magstep1

\centerline{{\bfv  $p$-ADIC AND ADELIC HARMONIC OSCILLATOR }}
\centerline{{\bfv WITH TIME-DEPENDENT FREQUENCY}}

\vskip1cm

\centerline{\bf Goran S. Djordjevi\'c$^{1}$ and Branko
Dragovich$^{2,3}$} 
\centerline{\it {${}^{1}$Department of Physics,
University of Ni\v s, P.O.Box 91, 18001 Ni\v s, Yugoslavia}}
\centerline{\it {${}^{2}$Institute of Physics,
P.O.Box 57, 11001 Belgrade, Yugoslavia}}
\centerline{\it {${}^{3}$Steklov Mathematical Institute,
Russian Academy of Sciences, Moscow, Russia}}
\vskip1cm

\noindent 
{\it The classical and quantum formalism for a $p$-adic and adelic
harmonic oscillator with time-dependent frequency is developed,
and general formulae for main theoretical quantities are obtained.
In particular, the $p$-adic propagator is calculated, and the
existence of a simple vacuum state as well as adelic quantum
dynamics is shown. Space discreteness and $p$-adic
quantum-mechanical phase are noted.}

\vskip1cm

\noindent {\bf{1. Introduction}}

\vskip.3cm

In quantum-mechanical experiments, as well as in all measurements,
numerical results belong to the field of rational numbers
$\msbm\hbox{Q}$. In principle, the corresponding theoretical
models could be made using only $\msbm\hbox{Q}$, but it would
missed usual effectiveness and beauty of mathematical analysis.
So, instead of $\msbm\hbox{Q}$ one traditionally applies the field
of real numbers $\msbm\hbox{R}$ in classical mechanics and the
field of complex numbers $\msbm\hbox{C}$ in quantum mechanics.
$\msbm\hbox{R}$ is completion of $\msbm\hbox{Q}$ with respect to
the metric induced by the absolute value and $\msbm\hbox{C}$ is an
algebraic extension of $\msbm\hbox{R}$. In addition to
$\msbm\hbox{R}$ there exist the fields of $p$-adic numbers 
$\msbm\hbox{Q}_p$ as completions of $\msbm\hbox{Q}$ with respect
to $p$-adic norms ($p$= a prime number) [1]. According to the
Ostrowski theorem, $\msbm\hbox{R}$ and $\msbm\hbox{Q}_p$ (for
every $p$) exhaust all possible completions of $\msbm\hbox{Q}$.
Thus $\msbm\hbox{Q}$ is dense not only in $\msbm\hbox{R}$ but also
in each $\msbm\hbox{Q}_p$. Therefore, in the last decade there has
been a lot and successful interest in construction of theoretical
models with $p$-adic numbers (for a review, see, Refs. 2-5).

There is a common belief that none separated prime number $p$
plays a special role in physics and that $p$-adic models have to
be taken together for all primes. It is clear that $p$-adic
models, having some  physical meaning,  must be somehow  connected
with the ordinary (real) ones. The space of adeles [6]  $\msbm\hbox{A}$ 
is a mathematical instrument which enables us to consider real
and $p$-adic numbers simultaneously and as a whole. Thus it is
natural to expect that adelic approach provides a more complete
description of a physical system than the ordinary one.

$p$-Adic numbers exhibit ultrametric (non-archimedean) properties,
which may be realized in quantum systems at very short distances.
Possibility that space-time at the Planck scale exhibits $p$-adic
and adelic structure is one of the main physical motivations to
investigate the corresponding models.

In order to start with a systematic approach to $p$-adic models of
quantum systems, $p$-adic quantum mechanics [7,8] was formulated.
Quantization is done along the Weyl procedure. The corresponding
Hilbert space $L_2(\msbm\hbox{Q}_p)$ contains complex-valued
square integrable functions on $\msbm\hbox{Q}_p$. Instead of the
Schr$\ddot{o}$dinger equation, the dynamical evolution and the
spectral problem of a system are related to the unitary
representation of the evolution operator $U_p(t)$ on
$L_2(\msbm\hbox{Q}_p)$. As a generalization and unification of
$p$-adic and ordinary quantum mechanics, recently was formulated
adelic quantum mechanics [9].

So far a rather small number of physical systems has been treated
in $p$-adic and adelic quantum mechanics: a non-relativistic free
particle [7], a harmonic oscillator [7,9], a particle in a
constant field [8],  the de Sitter minisuperspace model of the
universe [10] and a relativistic free particle [11]. It is
doubtless that evaluation of some other physical systems, which
exhibit $p$-adic and adelic properties, will give new insights
into this subject and new directions for future investigations
at the Planck scale.

In this paper we  show existence and some properties of $p$-adic
and  adelic harmonic oscillator with time-dependent frequency
(HOTDF). Model of the HOTDF has vast applications  from quantum
optics [12] to quantum cosmology [13]. Nevertheless, many properties of
classical and quantum motion can be found without specifying the
time dependence of $\omega(t)$.

\vskip1cm

\noindent
{{\bf 2.  $p$-Adic numbers and adeles}}

\vskip.3cm

To make this paper more self contained,  we give here a very
short review of some basic facts on $p$-adic numbers and adeles.

Any rational number $x\neq 0$ can be presented as $x=p^{\nu}{m\over n}$,
where $\nu, m, n\in \msbm\hbox{Z}$ and $p$ is a given prime number which
divides neither $m$ nor $n$. By definition, $p$-adic norm of $x$ is
$$
  \vert x\vert_p = p^{-\nu},\ \ \vert 0\vert_p = 0, \eqno(2.1)
$$
and holds the strong triangle inequality:
$$ \vert x+y\vert_p
\leq \hbox{max}(\vert x\vert_p,\vert y \vert_p ).  \eqno(2.2)
$$
A norm (valuation) with the property (2.2) is called
non-archimedean or ultrametric norm. Every $p$-adic number $x$ can
be uniquely presented by the canonical expansion 
$$ 
x =
p^{\nu}\sum_{i=0}^{+\infty} x_ip^i,\ \  x_i\in \{ 0,1,...,p-1\},
  \ \  x_0\neq 0, \ \ \nu \in \msbm\hbox{Z}            \eqno(2.3)
$$
The expansion (2.3) is convergent with
respect to the metric induced by $p$-adic norm, i.e.
$d_p(x,y)=\vert x -y\vert_p$.

There are mainly two kinds of analysis on $\msbm\hbox{Q}_p$ based
on two different maps: $\msbm\hbox{Q}_p \to \msbm\hbox{Q}_p $ and
$\msbm\hbox{Q}_p \to \msbm\hbox{C}$. We use both of these
analyses.

Elementary $p$-adic functions, like $\exp x$, $\sin x$ and $\cos
x$ are given by series of the same form as  in the real case.
However, the region of convergence is rather restricted and it is
$\vert x\vert_p < \vert 2\vert_p $ for the above functions.
Derivatives of $p$-adic valued functions are defined as in the
real case, but using $p$-adic norm instead of the absolute value.

For complex-valued functions of $p$-adic argument there is
well-defined integration with the Haar measure. In particular, we
use the Gauss integral [3]  
$$ 
\int_{\mid x\mid_{p}\le
p^{\nu}}\chi_p(\alpha x^2+\beta x)dx =
\cases{p^\nu\Omega(p^\nu\vert\beta\vert_p),&$\vert\alpha\vert_p\le
p^{-2\nu},$\cr \lambda_p(\alpha)\vert 2\alpha\vert_p^{-1/2}
\chi_p\big(-{\beta^2\over4\alpha}\big)\Omega\big(p^{-\nu}\vert{\beta\over2
\alpha} \vert_p\big),& $\vert
4\alpha\vert_p>p^{-2\nu}.$\cr}\eqno(2.4) 
$$ 
$\chi_p(u) = \exp(2\pi
i\{u\}_p)$ is a p-adic additive character, where $\{u\}_p$ denotes
the fractional part of $u\in\msbm\hbox{Q}_p$. $\lambda_p(\alpha)$
is an arithmetic complex-valued function  with the following basic
properties [3]: 
$$ 
\lambda_p(0) = 1, \ \lambda_p(a^2\alpha) =
\lambda_p(\alpha), \ \lambda_p(\alpha)\lambda_p(\beta) =
\lambda_p(\alpha+\beta)\lambda_p(\alpha^{-1}+\beta^{-1}), \
\vert\lambda_p(\alpha)\vert_\infty = 1.\eqno(2.5) 
$$ 
$\Omega(\vert
u\vert_p)$ is the characteristic function on $\msbm\hbox{Z}_p$,
{\it i.e.} 
$$ 
\Omega(\vert u\vert_p) = \cases{1,&$\vert
u\vert_p\le1,$\cr 0,&$\vert u\vert_p>1,$\cr}\eqno(2.6) 
$$ 
where
$\msbm\hbox{Z}_p = \{x\in \msbm\hbox{Q}_p: \vert x\vert_p\le1\}$
is the ring of p-adic integers.

An adele [6] $a\in\msbm\hbox{A}$ is an infinite sequence 
$$ 
a =
(a_\infty,a_2,\cdots,a_p,\cdots)\ ,\eqno(2.7) 
$$ 
where
$a_\infty\in\msbm\hbox{R}$ and $a_p\in\msbm\hbox{Q}_p$ with the
restriction that $a_p\in\msbm\hbox{Z}_p$ for all but a finite set
$S$ of primes $p$. The set of all adeles $\msbm\hbox{A}$ can be
written in the form 
$$ 
\msbm\hbox{A} = \mathop{U}\limits_{S}{\cal
A}(S), \quad {\cal A}(S) = \msbm\hbox{R}\times \prod_{p\in
S}\msbm\hbox{Q}_p\times\prod_{p\not\in S}\msbm\hbox{Z}_p\
.\eqno(2.8) 
$$ 
$\msbm\hbox{A}$ is a topological space. It is a
ring with respect to componentwise addition and multiplication.
There is a natural generalization of analysis on $\msbm\hbox{R}$
and $\msbm\hbox{Q}_p$ to analysis on $\msbm\hbox{A}$.

\vskip1cm

\noindent
{{\bf 3.  Classical oscillator: real, $p$-adic and
adelic case}}

\vskip.3cm

Classical HOTDF is given by the Lagrangian 
$$ 
L(x,\dot{x},t) =
{{m}\over{2}}\dot{x}^2 - {{m\omega^2(t)}\over{2}}x^2\ , \eqno(3.1)
$$ 
where $m\in \msbm\hbox{Q} $. Time-dependent frequency $\omega(t) = 
\sum_{n\geq 0}\omega_n t^n, $ where $\omega_n \in \msbm\hbox{Q}$,
is assumed to be an analytic function on $\msbm\hbox{D}_{\infty}
\subset \msbm\hbox{R}$ and on
$\msbm\hbox{D}_p \subset \msbm\hbox{Z}_p$ for all $p$.  In other words,
when $t\in {\cal A}(S)$ then $\omega (t) \in {\cal A}(S^{\prime})$,  
where $S$ and $S^{\prime}$ are some finite sets of primes $p$.
In the real case
$m,x,\dot x,t,\omega(t)\in
\msbm\hbox{R}\equiv\msbm\hbox{Q}_\infty$ (in the sequel index
$\infty$ denotes quantities defined on $\msbm\hbox{R}$ or
$\msbm\hbox{C}$) and the analogous situation is for the $p$-adic
counterparts. Because of formal similarity of analyses, evaluation
of  (3.1) is the same in real and $p$-adic dynamics. Thus, in real
and $p$-adic  cases, the equation of motion is  
$$ 
\ddot{x}(t) +\omega^2(t)x(t) = 0 \eqno(3.2) 
$$ 
with general solution [14]
$$
x(t) = G(t)[C_1 \cos\gamma(t) + C_2 \sin\gamma(t)]. \eqno(3.3)
$$
The amplitude $G(t)$ and phase $\gamma(t)$ satisfy equations 
$$
G^3(t)\ddot{G}(t) + \omega^2(t)G^4(t) = C^2, \ \ \dot{\gamma}(t)
G^2(t)=C, \eqno(3.4) 
$$ 
where $C$ is a constant $(0< C\in \msbm\hbox{R}, \ \ C \in 1 + p\msbm\hbox
{Z}_p)$ and can be taken $C = 1$. We are interested in analytic solution
of (3.3), where $G(t)$ and $\gamma (t)$ are power series in $t$
with rational coefficients. Differential equation for $G(t)$ is non-linear.
However, it does not lead to non-linear algebraic equations
for unknown coefficients $G_n$ in expansion $G(t)=\sum_{n\geq 0} G_n t^n$
and any $G_n$ can be presented as a rational number, which is the same
in the real and all $p$-adic cases. 
Note that usual power series with rational coefficients which are 
convergent on 
$\msbm\hbox{D}_{\infty}\subset \msbm\hbox{R}$ in the real case 
are also $p$-adically convergent
in some region $\msbm\hbox{D}_p \subset \msbm\hbox{Z}_p $. 

As an illustration of analytic solutions of the equations (3.4)
(with $C=1$) we present two simple examples.

\noindent {\it Example 1.}  Let $\omega (t) = \omega_0 / (1+at)^2$,
where $\omega_0 = b^{-2}$ and $a,b\in \msbm\hbox{N} $.
Then
$$
G(t) = b(1 + at), \  \  \  \  \ \gamma (t) = {{1}\over{b^2}}
{{t}\over{1+at}} .
$$
Since  $\gamma (t)$ is argument of trigonometric
functions in (3.3) one obtains that common region of convergence
for all analytic expansions is $\mid t\mid_p < \mid 2b^2 \mid_p  $
for each $p$.

\noindent {\it Example 2.}  Let $\omega (t)= \omega_0 /(1+at)$,
where $\omega_0 = b^{-2}(1+a^2 b^4 /4)^{{{1}\over{2}}}$
and $a,b\in 2\msbm\hbox{N}$. Then in an analogous way to the Example 1
we get:
$$
G(t)=b(1+at)^{{{1}\over{2}}} , \ \ \  \gamma (t)={{1}\over{ab^2}}\ln (1+at)
, \ \ \ \mid t\mid_p < \mid 2b^2 \mid_p  .
$$
Thus,
there exist non-trivial adelic solutions for $G(t)$, and $\gamma (t)$,
and conseqently for $x(t)$ in the form (3.3).

To determine constants $C_1$  and  $C_2$ we use two kinds of conditions
on the classical trajectory.

\vskip.5cm 

\noindent {\it {\bf (3.1) Solution with the end point
conditions}} 

\vskip.3cm

The classical trajectory that links two space-time points
$(x',t')$ and $(x'',t'')$ is  
$$
x(t)={{G(t)}\over{\sin(\gamma''-\gamma')}}\bigg[
{{x'}\over{G'}}\sin(\gamma''-\gamma(t))+{{x''}\over{G''}}
\sin(\gamma(t)-\gamma') \bigg] ,  \eqno(3.5)
$$ 
where $x'=x(t'), \
x''=x(t''),\  G'=G(t'),\  G''=G(t''),\  \gamma'=\gamma(t')$ and
$\gamma''=\gamma(t'')$. Note that condition $  \gamma''-\gamma'
\neq m\pi ,\ m\in \msbm\hbox{Z}$,  must be satisfied in the real
case. Recall also that $\vert \gamma'' - \gamma'\vert_p < \vert
2\vert_p $. As we shall see later it is useful to write the
corresponding momentum in the form 
$$ 
k(t)=m\dot{x}(t) = m
{{\dot{G}(t)}\over{G(t)}}x(t) + {{m
G(t)\dot{\gamma}(t)}\over{\sin(\gamma''-\gamma')}}\bigg[
{{x''}\over{G''}}
\cos(\gamma(t)-\gamma')-{{x'}\over{G'}}\cos(\gamma''-\gamma(t))
\bigg].   \eqno(3.6) 
$$

\vskip.5cm 

\noindent   {\it {\bf (3.2) Solution with the initial
conditions}} 

\vskip.3cm

Imposing the initial conditions $x^0=x(t^0),\ k^0=m\dot{x}(t^0)$
we find evolution of the classical state as follows: 
$$ 
x(t) =
\bigg[ {{G(t)}\over{G^0} } \cos(\gamma(t)-\gamma^0)
-{{G(t)\dot{G}^0}\over{C}} \sin(\gamma(t)-\gamma^0)\bigg]x^0 +
{{G(t)G^0}\over{m C}}\sin(\gamma(t)-\gamma^0) k^0 \ ,   
$$ 
$$ 
k(t)
= \bigg[m\bigg({{\dot{G}(t)}\over{G^0}} -
{{G(t)\dot{\gamma}(t)\dot{G}^0}\over{C}}
\bigg)\cos(\gamma(t)-\gamma^0)  -m
\bigg({{\dot{G}(t)\dot{G}^0}\over{C}}
+{{G(t)\dot{\gamma}(t)}\over{G^0}} \bigg)\sin(\gamma(t)-\gamma^0)
\bigg]x^0 $$    $$  + {{G^0}\over{C}}\big[
G(t)\dot{\gamma}(t)\cos(\gamma(t)-\gamma^0) + \dot{G}(t)
\sin(\gamma(t)-\gamma^0) \big] k^0 ,    \eqno(3.7)
$$ 
where $ G^0 =
G(t^0)$ and $\gamma^0 = \gamma(t^0)$.  Putting first $t=t'$ and
then $t=t''$ in the first equation of (3.7) one can find $x^0$ and
$k^0$ as functions of $x'$ and $x''$. Inserting these
$x^0=x^0(x',x'')$ and $k^0=k^0(x',x'')$ into the second equation
of (3.7) one gets the same formula (3.6) for $k(t)$.

A suitable way to calculate the corresponding classical action 
$$
\bar{S}(x'',t'';x',t') =
{{m}\over{2}}\int_{t'}^{t''}[\dot{x}^2(t)-\omega^2(t)x^2(t)]dt
\eqno(3.8) 
$$ 
is integrating by parts and using  the equation of
motion (3.2). It leads to 
$$ 
\bar{S}(x'',t'';x',t') =
{{m}\over{2}}(x''\dot{x}'' -x'\dot{x}'). \eqno(3.9) 
$$ 
In virtue
of (3.6), that gives $\dot{x}$ as function of $x$, we find action
in the form quadratic in $x''$ and $x'$, {\it i.e.}
$$
\bar{S}(x'',t'';x',t') = {{m}\over{2}}\bigg[\bigg(
{{\dot{\gamma}''}\over{\tan(\gamma''-\gamma')}} +
{{\dot{G}''}\over{G''}} \bigg)x''^2 $$    $$
-{{2\sqrt{\dot{\gamma}''\dot{\gamma}'}}\over{\sin(\gamma''-\gamma')}}
x''x' + \bigg(
{{\dot{\gamma}'}\over{\tan(\gamma''-\gamma')}}-{{\dot{G}'}\over{G'}}
\bigg) x'^2 \bigg],       \eqno(3.10)
$$ 
where we used equality 
$$
{{G''\dot{\gamma''}}\over{G'}}+{{G'\dot{\gamma'}}\over{G''}}=
2\sqrt{\dot{\gamma''}\dot{\gamma'}}\ , 
$$ 
which is derived by
means of (3.4).

Note that the above $p$-adic formalism has the same form as its
real counterpart, or in other words, the classical HOTDF is
invariant under change of the number field $\msbm\hbox{R}$ and
$\msbm\hbox{Q}_p$, for every $p$. This may be regarded as a
necessary condition for existence of an adelic classical HOTDF,
which we construct in the following way. Let the position $x$,
momentum $k$ and time $t$ be adelic quantities, like (2.7). The
corresponding adelic Langrangian is given by  
$$
L(x,\dot{x},t) =
\big( L(x_\infty,\dot{x}_\infty,t_\infty),\ L(x_2,\dot{x}_2,t_2),\
..., \ L(x_p,\dot{x}_p,t_p),\ ... \big),  
$$ 
where
$L(x_v,\dot{x}_v,t_v)= m[\dot{x}^2_v-\omega^2(t_v)x^2_v]/2 $ with
$v=\infty,2,...,p,...$, and $\vert L(x_p,\dot{x}_p,t_p)\vert_p
\leq 1$ for all but a finite number of primes $p$. Also, all the
other above introduced quantities, regarded as real and $p$-adic,
can be generalized to the adelic ones. For instance, adelic
classical action is 
$$
\bar{S}(x'',t''; x',t')= \big(
\bar{S}(x''_\infty,t''_\infty;x'_\infty,t'_\infty),\
 \bar{S}(x''_2,t''_2;x'_2,t'_2),\
...,\ \bar{S}(x''_p,t''_p;x'_p,t'_p),\ ... \big)\ ,   \eqno(3.11)
$$ 
where real and $p$-adic ingredients have the form (3.10).

\vskip1cm

\noindent
{{\bf 4.  Quantum oscillator: real, $p$-adic and adelic case}}

\vskip.5cm

The unique formalism of ordinary quantum mechanics which enables
$p$-adic and adelic generalization with complex-valued wave
functions is a triple [7,9]  
$$
\big( L_2(\msbm\hbox{R}),\ W(z_\infty),\
U(t_\infty) \big)\ ,    \eqno(4.1) 
$$   
where $L_2(\msbm\hbox{R})$
is the Hilbert space, $z_\infty$ is a point of real classical
phase space, $W(z_\infty)$ is a unitary representation of the
Heisenberg-Weyl group on $L_2(\msbm\hbox{R})$, and $U(t_\infty)$
is a unitary representation of the evolution operator on $
L_2(\msbm\hbox{R})$. Hence, under $p$-adic and adelic
quantum mechanics we  understand $p$-adic and adelic analogues of
(4.1), {\it i.e.} 
$$ 
\big( L_2(\msbm\hbox{Q}_p),\ W(z_p),\ U(t_p)
\big)\ ,     \eqno(4.2) 
$$   
$$
\big( L_2(\msbm\hbox{A}),\ W(z),\
U(t) \big)\ ,    \eqno(4.3) 
$$ 
respectively.  Thus, to find adelic
eigenstate and  its evolution of a HOTDF given by $U(t'',t')$, one
has to solve the equation 
$$ U(t'',t')\Psi_S^{(\alpha)}(x',t')
=\chi[\alpha(\gamma''-\gamma')]\Psi_S^{(\alpha)}(x',t')\ ,
\eqno(4.4) 
$$ 
where $\alpha = (\alpha_\infty, \alpha_2, ...,
\alpha_p, ...)$ is an adelic analogue of energy,
$\chi(u)=\prod_v\chi_v(u_v)=\exp{(-2\pi i u_\infty )}
\prod_p \exp{(2\pi i \{ u_p\}{_p})}$ and
$$
\Psi_S^{(\alpha)}(x,t)=\Psi^{(\alpha_\infty)}_\infty(x_\infty,t_\infty)
\prod_{p\in S}\Psi^{(\alpha_p)}_p(x_p,t_p)\prod_{p\not\in
S}\Omega(\mid x_p \mid_p)\ . \eqno(4.5) 
$$ 
The evolution operator
$U(t'',t') = \prod_v U_v(t''_v,t'_v)$ acts componentwise as
follows: 
$$
[U_v\Psi_v](x''_v,t''_v) = \int_{\msbm\hbox{Q}_v}
{\cal K}_v (x''_v,t''_v;x'_v,t'_v) \Psi_v(x'_v,t'_v)dx'_v\ .
\eqno(4.6) 
$$ 
The kernel $ {\cal K}_v(x''_v,t''_v;x'_v,t'_v))  $
is defined by the Feynman path integral 
$$
{\cal K}_v
(x''_v,t''_v;x'_v,t'_v)= \int\chi_v\big(-{{1}\over{h}}S[x] \big)
{\cal D}x = \int\chi_v \big( -{{1}\over{h}} \int_{t'_v}^{t''_v}
L(x_v,\dot{x}_v,t_v)dt_v \big) \prod_{t_v}dx(t_v)\ ,    \eqno(4.7)
$$ 
where $h$ is the Planck constant.  The kernel ${\cal K}$, also
called the quantum-mechanical propagator, is of central importance
not only in ordinary but also in $p$-adic and adelic quantum
mechanics.

The $p$-adic Feynman path integral for classical actions quadratic
in $x''$ and $x'$  is calculated in [15]  and has the same
form as its real counterpart. Namely, if $
\bar{S}(x''_v,t''_v;x'_v,t'_v) $ is quadratic in $x''_v$ and
$x'_v$ then  
$$
{\cal K}_v(x''_v,t''_v;x'_v,t'_v) = \lambda_v\bigg(
-{{1}\over{2h}} {\partial^2\bar{S}\over{\partial x''_v\partial
x'_v }}\bigg) \bigg |
{{1}\over{h}}{{\partial^2\bar{S}\over{\partial x''_v\partial
x'_v}}} \bigg |_v^{1/2} \chi_v \bigg(-{{1}\over{h}}
\bar{S}(x''_v,t''_v;x'_v,t'_v)\ \bigg)\ , \eqno(4.8)
$$ 
where $
\lambda_\infty(\alpha) = (1-i\ sign\ \alpha)/\sqrt{2}$\ ,
$\lambda_\infty (0) = 1 $ and  satisfies properties (2.5).

Applying formula (4.8) to the HOTDF  and using (3.10), we get
$$
{\cal K}_v(x''_v,t''_v;x'_v,t'_v) = \lambda_v\bigg(
{{m}\over{2h}}{{\sqrt{\dot{\gamma}''\dot{\gamma}'}\over{\sin
(\gamma'' - \gamma')}}} \bigg) \bigg |
{{m}\over{h}}{{\sqrt{\dot{\gamma}''\dot{\gamma}'}\over{\sin
(\gamma'' - \gamma')}}} \bigg|_v^{1/2}$$  $$ \chi_v\bigg \{
-{{m}\over{2h}}\bigg[\bigg(
{{\dot{\gamma}''}\over{\tan(\gamma''-\gamma')}} +
{{\dot{G}''}\over{G''}} \bigg)x''^2
-{{2\sqrt{\dot{\gamma}''\dot{\gamma}'}}\over{\sin(\gamma''-\gamma')}}
x''x' + \bigg(
{{\dot{\gamma}'}\over{\tan(\gamma''-\gamma')}}-{{\dot{G}'}\over{G'}}
\bigg) x'^2 \bigg]\bigg \}\ ,  \eqno(4.9)
$$ 
that contains the
earlier obtained result in the real case (see, e.g. Ref. 14). One
can explicitly show that the propagator (4.9) satisfies all usual
properties of the probability amplitude for a quantum particle to
go from a space-time point $(x'_v,t'_v)$ to a space-time point
$(x''_v,t''_v)$.

The corresponding adelic propagator is  $$    
{\cal
K}(x'',t'';x',t') = \prod_v {\cal K}_v (x''_v,t''_v;x'_v,t'_v)\ ,
\eqno(4.10) 
$$ 
where ${\cal K}_v (x''_v,t''_v;x'_v,t'_v)  $ is
given by (4.9). Product in (4.10) is divergent, but it may be
regarded as an adelic functional on the space of test functions
which are the adelic Schwartz-Bruhat functions   (see, also [16]).

In $p$-adic  quantum mechanics a significant role plays the
eigenstate $\Omega (\vert x_p \vert_p)$ (2.6), which is invariant
under $U_p(t''_p,t'_p)$ transformation and may be regarded as
$p$-adic vacuum state since it has $\{ \alpha_p(\gamma''_p -
\gamma'_p)\}_p = 0$. Due to (4.5), the existence of $\Omega(\vert
x_p\vert_p)$ for all but a finite number of $p$ is a necessary
condition for unification of ordinary and $p$-adic quantum
mechanics in the form of adelic one. This $\Omega$-state exists
iff 
$$
\int_{\mid x'_p \mid_p \leq 1} {\cal K}_p
(x''_p,t''_p;x'_p,t'_p) dx'_p = \Omega(\vert x''_p \vert_p )
\eqno(4.11) 
$$ 
is satisfied.

Inserting (4.9) with $v=p$ into (4.11), and using the integral
(2.4) for $\nu =0$, we can derive some conditions on $G(t_p),\
\gamma(t_p)$ and $m$, which provide $\Omega$-eigenstate. For
example, if $\gamma(t_p) = \gamma_0 + \gamma_1t_p + \gamma _2t_p^2
+ ... + \gamma_nt_p^n + ....$ and  
$$
\bigg |
{\dot{G}'\over{G}'}\bigg |_p < \bigg
|{{\dot{\gamma}'}\over{\tan(\gamma'' - \gamma')}}\bigg |_p >
\bigg| {{h}\over{2m}}\bigg|_p
$$
then $\Omega(\vert x_p \vert_p)$ exists
for all $p\neq 2$. It is worth noting  that  not every HOTDF
has $\Omega$-state and may be adelically generalized.

As the simplest illustration of the above expressions one can take
frequency $\omega(t) = \omega_0 $ and recover earlier obtained
result [9].

\vskip1cm

\noindent
{{\bf 5.  Concluding remarks}}

\vskip.5cm

According to (4.5) adelic wave function $\Psi(x,t)$ offers more
information on a physical system than only its standard part
$\Psi_\infty(x_\infty,t_\infty)$. Let us note here space
discreteness and $p$-adic phase, which are generic and mainly
follow from adelic quantum formalism.

For example,   adelic state 
$$ 
\Psi(x,t) =
\Psi_\infty(x_\infty,t_\infty)\prod_p \Omega(\vert x_p \vert_p) 
$$
exhibits discrete structure of the space at the length $l_0 =
(hm^{-1}\omega^{-1})^{1/2}$. Namely, according to the usual
interpretation of the wave function we have to consider $\vert
\Psi(x,t)\vert^2_\infty $ at rational points $x$ and $t$.  In the
above adelic case we get 
$$
\vert \Psi(x,t)\vert^2_\infty =
\vert\Psi_\infty(x,t)\vert^2_\infty \prod_p \Omega(\vert x\vert_p)
= \cases{ \vert \Psi_\infty(x,t)\vert^2_\infty\ , & $
x\in\msbm\hbox{Z}\ , $\cr 0\ , & $ x\in \msbm\hbox{Q}\setminus
\msbm\hbox{Z}\ . $\cr}  
$$ 
Here we used the following properties
of $\Omega$-function: $\Omega^2(\vert x\vert_p) = \Omega(\vert
x\vert_p),\ \prod_p\Omega(\vert x\vert_p) =1$ if
$x\in\msbm\hbox{Z}$, and $\prod_p\Omega(\vert x\vert_p) =0$ if
$x\in\msbm\hbox{Q}\setminus\msbm\hbox{Z}$. Thus, it means that
position $x$ may have only discrete values: $x/l_0 =0,\pm 1,\pm
2,...$ To verify this space discreteness experimentally one has to
examine physical system in its vacuum state and at distances
characterized by the length $l_0$. When system is in a rather
mixed state 
$$ 
\Psi(x,t) =\sum_{S,\alpha} C(S,\alpha)
\Psi^{(\alpha)}_S(x,t) 
$$ 
the sharpness of the discrete structure
disappears and space demonstrates usual continuous properties.
So, this space discreteness is a quantum effect and depends on
adelic quantum state.

Adelic wave function gives also a framework to investigate a new
kind of phase, which may be called $p$-adic phase. In fact, (4.5)
contains 
$$
\Psi_p(x_p,t_p) =
\chi_p[\alpha_p(\gamma(t_p)-\gamma^0)]\Psi_p(x_p,0)\ , 
$$ 
where
$\chi_p[\alpha_p(\gamma(t_p) -\gamma^0)]$ presents a $p$-adic
dynamical phase.  This may be observed investigating the fine
structure of interference phenomena.

At real distances which are very large in comparison with $l_0$,
$p$-adic effects become hidden and adelic quantum mechanics
reduces to  the ordinary one. In such case we have to integrate
$\vert\Psi (x,t)\vert^2$ over $p$-adic components of adelic space.
Since $\int_{\vert x_p\vert_p\leq1} dx = 1$ and
$\int_{\msbm\hbox{Q}_p}\vert \Psi_p(x_p,t_p)\vert^2_\infty dx_p =
1$ we have 
$$ 
\int_{{\cal A}(S)\setminus\msbm\hbox{R}} \vert
\Psi_S(x,t)\vert^2_\infty dx = \vert \Psi_\infty
(x_\infty,t_\infty)\vert^2_\infty dx_\infty \prod_{p\in
S}\int_{\msbm\hbox{Q}_p} \vert \Psi_p(x_p,t_p)\vert^2_\infty dx_p
$$  $$\times \prod_{p \notin S}\int_{\msbm\hbox{Z}_p} \Omega(\vert
x_p \vert_p) dx_p   =\vert\Psi_\infty(x_\infty,t_\infty
\vert^2_\infty dx_\infty\ . 
$$ 
Hence, ordinary quantum theory may
be regarded as an effective approximation of the more profound
adelic one.

 \vskip1cm

\noindent
{\bf{ Acknowledgements}}

\vskip.5cm

This work is partially supported by the Russian Foundation for Basic
Research, Grant No 990100166, and leading scientific schools. 
We wish to thank colleagues
from the Department of Mathematical Physics of the Steklov
Mathematical Institute in Moscow for fruitful discussions.

\vskip1cm

\noindent
{{\bf REFERENCES}}

\vskip.5cm
\item{1.} W.H. Schikhof, {\it Ultrametric Calculus}, Cambridge Univ. Press,
Cambridge (1984).
\item{2.} L. Brekke and P.G.O. Freund, {\it Phys. Rep.} {\bf 233}, 1 (1993).
\item{3.} V.S. Vladimirov, I.V.Volovich and E.I. Zelenov,
{\it $p$-Adic Analysis and Mathematical Physics}, World Scientific,
Singapore (1994).
\item{4.} A. Khrennikov, {\it $p$-Adic Valued Distributions in
Mathematical Physics}, Kluwer Acad. Publishers, Dordrecht (1994).
\item{5.} A. Khrennikov, {\it Non-Archimedean Analysis: Quantum Paradoxes,
Dynamical Systems and Biological Models}, Kluwer Acad. Publishers,
Dordrecht (1997).
\item{6.} I.M. Gel'fand, M.I.Graev and I.I. Piatetskii-Shapiro,
{\it Representation Theory and Automorphic Functions}, Nauka, Moscow,
(1966);

A. Weil, {\it Adeles and Algebraic Groups}, Birkh$\ddot{a}$user, Boston,
(1982).
\item{7.} V.S. Vladimirov and I.V. Volovich, {\it Commun. Math. Phys.}
{\bf 123}, 659 (1989).
\item{8.} Ph. Ruelle, E. Thiran, D. Verstegen and J. Weyers,
{\it J. Math. Phys.} {\bf 30}, 2854 (1989).
\item{9.} B. Dragovich, {\it Theor. Math. Phys.} {\bf 101}, 1404 (1994);
{\it Int. J. Mod. Phys.} {\bf A 10}, 2349 (1995).
\item{10.} B. Dragovich, {\it Adelic Wave Function of the Universe},
in: Proceedings of the Third A. Friedmann International Seminar on
Gravitation and Cosmology, Friedmann Laboratory Publishing,
St.Petersburg (1995).
\item{11.} G.S. Djordjevi\'c, B. Dragovich and Lj. Ne\v si\'c,
{\it Mod. Phys. Lett.} {\bf A 14}, 317 (1999).
\item{12.} A.D. Jannussis and B.S. Bartzis, {\it Phys. Lett.} {\bf A 129},
263 (1988);

R.J. Glauber, {\it Quantum theory of particle trapping by oscillating
fields}, in: Quantum

Measurements in Optics, Plenum Press, New York (1992);

V.V. Dodonov, O.V. Man'ko and V.I. Man'ko, {\it Phys. Lett.} {\bf A 175},
1 (1993).
\item{13.} J.J. Halliwell and S.W. Hawking, {\it Phys. Rev.} {\bf D 31},
1777 (1985).

\item{14.} D.C. Khandekar and S.V. Lawande, {\it J. Math. Phys.} {\bf 16},
384 (1975);

C.P. Natividade, {\it Am. J. Phys.} {\bf 56}, 921 (1988);

C. Farina and A.J. Segui-Santonja, {\it Phys. Lett.} {\bf A 184}, 23 (1993).

\item{15.} G.S. Djordjevi\'c and B. Dragovich, {\it Mod. Phys. Lett.} 
{\bf A 12}, 1455 (1997).

\item{16.} B. Dragovich, {\it Integral Transforms and Special Functions}
{\bf 6}, 197 (1998).
\end